# Quo Vadis? Scientific Discovery in the Age of Artificial Intelligence

Petr O. Jedlička[1]

**Abstract**

*This paper examines the growing role of AI in scientific discovery. It first surveys the rapid rise of AI capabilities, especially in reasoning, abstraction, planning, and long-horizon task execution, before turning to scientometric evidence of AI's diffusion across the sciences. It then proposes a typology of AI systems used in research, ranging from specialized scientific AI through scientific AI assistants and agents to hybrid experimental systems that combine computation and physical experimentation. On this basis, it offers a selective overview of recent achievements in mathematics and computer science, physics, chemistry, the life sciences, and the behavioural and social sciences. It argues that, despite these advances, current systems remain constrained by important technical, epistemic, and institutional limitations, and that their growing use introduces both near-term and longer-term risks. The conclusion further suggests that the advancement of AI in science raises broader questions concerning the division of cognitive labour between human researchers and machines.*

**Keywords:** Artificial Intelligence, Scientific Discovery, AI Scientists, Research Automation, Human–AI Collaboration

---

[1] Institute of Philosophy, Czech Academy of Sciences, Jilská 361/1, 110 00 Prague 1, Czechia, jedlicka@flu.cas.cz

## 1. Introduction

The public release of ChatGPT in late 2022 marked a turning point in the relationship between artificial intelligence and scientific research. Although AI had already been employed extensively across a variety of scientific domains, its applications were largely confined to specialised machine-learning (ML) systems designed for narrowly defined tasks such as classification, prediction, optimisation, and diagnostics. The emergence of large foundation models and their widespread public availability dramatically altered this situation. For the first time, highly capable AI systems became directly accessible to researchers across disciplines, contributing to a rapid integration of AI into scientific practice.

From a historical perspective, these developments represent the latest stage in a much longer effort to employ computational systems in scientific discovery. Early attempts can be traced to symbolic computer-assisted discovery systems developed during the 1970s and 1980s – for instance BACON, which attempted to discover simple numerical laws relating two variables by applying heuristic search strategies (Langley et al. 1987)[2]. Such systems, however, were constrained by the limitations of symbolic approaches characteristic of classical computational methods and remained limited in their capacity for autonomous scientific discovery. The symbolic systems nevertheless provided substantial assistance to scientists by supplying the computational infrastructure required for experimentation, data processing, and the analysis of massive datasets in data-intensive disciplines such as particle physics, astronomy, and genetics. These systems remained auxiliary in the sense that all of their underlying logic and rules were explicitly formulated by human scientists.

Subsequent advances in probabilistic algorithms and ML gradually expanded the role of AI within science. Beginning in the 1990s, ML techniques increasingly appeared in a number of scientific domains. Nevertheless, these systems typically remained highly specialised. Their scientific usefulness depended upon extensive training, and they rarely generalised beyond the domain-specific problems for which they had been designed.

A major advance occurred with the introduction of transformer architectures in 2017. Unlike earlier approaches, transformer-based systems demonstrated an unprecedented capacity for abstraction, generalisation, reasoning, and knowledge integration across domains. These developments laid the foundations for contemporary large language models (LLMs) and other foundation models, which increasingly exhibit capabilities of generalized intelligence. As a result, AI has evolved from a collection of narrowly specialised computational tools into a broadly applicable technology that now influences virtually every scientific discipline.

[2] Pat Langley, Herbert A. Simon, and Gary L. Bradshaw, "Heuristics for Empirical Discovery," in Computational Models of Learning, Symbolic Computation. ed. Leonard Bolc (Berlin and Heidelberg: Springer-Verlag, 1987), 21–54. https://doi.org/10.1007/978-3-642-82742-6_2

The central claim of this paper is that AI has already become a significant participant in scientific research and discovery. Unlike earlier symbolic and ML systems, contemporary frontier AI systems are capable of performing highly autonomous research activities that closely mirror processes previously carried out exclusively by human scientists, particularly when consisting of several specialised agents that perform different roles and share the cognitive labour involved in scientific tasks. In such a research setting, the role of the human scientist can be reduced largely to defining the research objective, prompting the AI system, and supervising its operation. Human researchers no longer necessarily provide the primary knowledge and creative contributions that characterised earlier discoveries. This represents a genuinely new situation in the history of scientific research.

Contemporary AI systems have already demonstrated capacities that only recently appeared unattainable – from theorem proving and algorithmic discovery to protein structure prediction. Yet these successes coexist with several shortcomings, including hallucinations, epistemic opacity, limited robustness, and the necessity of human oversight. Both achievements and limitations of current AI systems suggest a considerably more complex picture than their assessments typically acknowledge. Beyond questions of technical performance, the growing role of AI in science raises a series of questions about their epistemic and social impact on the sciences as well as society. As AI systems assume increasingly active roles within research processes, traditional distinctions between tool, collaborator, and a genuine discoverer become progressively more difficult to sustain. The broader implications of these developments constitute the focus of the concluding sections of this article.

## 2. The Rise of AI Capabilities

The growing usefulness of AI for scientific research is closely connected to the rapid improvement of model capabilities during the past several years. Contemporary LLMs increasingly demonstrate abilities that include abstraction, reasoning, planning, and problem solving (Ke et al. 2025)[3], although important limitations remain (Lee et al. 2025)[4]. These advances have substantially expanded the range of scientific tasks that AI systems can perform and have increased their relevance across scientific disciplines.

One indication of this trend is the rapid improvement of model performance on benchmarks. In many domains, AI systems have progressed from performance levels comparable to secondary education toward results approaching, and occasionally exceeding, expert-level competence. This development has repeatedly rendered existing benchmarks obsolete, necessitating the creation of more demanding evaluations capable of distinguishing among frontier models (Center for AI Safety, Scale AI, and HLE Contributors Consortium 2026).[5] The timeline of benchmark evolution itself has therefore become an indirect indicator of accelerating AI capabilities.

A notable example is *Humanity's Last Exam* (HLE), a benchmark consisting of 2,500 questions situated at the frontier of human expertise and designed to assess performance across mathematics, the natural sciences, engineering, and the humanities. Early evaluations conducted in 2025 showed that frontier models performed poorly on these tasks, frequently achieving accuracies below 10 per cent. Within a remarkably short period, however, newer generations of models demonstrated substantial improvements. Frontier systems such as Gemini 3.1 Pro, and GPT-5.5 subsequently exceeded 40 per cent accuracy on the benchmark (Center for AI Safety 2026).[6] Although these results still remain below the level of human

---

[3] Zixuan Ke, Fangkai Jiao, Yifei Ming, Xuan-Phi Nguyen, Austin Xu, Do Xuan Long, Minzhi Li, Chengwei Qin, Peifeng Wang, Silvio Savarese, Caiming Xiong, and Shafiq Joty, "A Survey of Frontiers in LLM Reasoning: Inference Scaling, Learning to Reason, and Agentic Systems," arXiv, 2025, https://doi.org/10.48550/arXiv.2504.09037.

[4] Seungpil Lee, Woochang Sim, Donghyeon Shin, Wongyu Seo, Jiwon Park, Seokki Lee, Sanha Hwang, Sejin Kim, and Sundong Kim, "Reasoning Abilities of Large Language Models: In-Depth Analysis on the Abstraction and Reasoning Corpus," *ACM Transactions on Intelligent Systems and Technology* 16, no. 6 (2025): Article 137, 1–52, https://doi.org/10.1145/3712701.

[5] Center for AI Safety, Scale AI, and HLE Contributors Consortium, "A Benchmark of Expert-Level Academic Questions to Assess AI Capabilities," *Nature* 649 (2026): 1139–1146, https://doi.org/10.1038/s41586-025-09962-4.

[6] Center for AI Safety, *Humanity's Last Exam*, accessed June 12, 2026, https://lastexam.ai/.

experts, they nevertheless illustrate the extraordinary pace at which capabilities have improved.

A particularly important example is also provided by the Abstraction and Reasoning Corpus (ARC-AGI), which was designed to evaluate forms of general intelligence associated with learning new skills and solving novel problems – that is, tasks that are often described as "easy for humans, difficult for AI." Unlike conventional benchmarks, ARC tasks require the discovery of previously unseen rules and therefore provide a useful measure of generalisation and adaptive reasoning. The more recent ARC-AGI-2 and ARC-AGI-3 benchmarks further increased task complexity and has become an important indicator of progress in machine reasoning (Chollet et al. 2026)[7].

Results from the ARC Prize competition illustrate both the rapid progress and the remaining limitations of contemporary systems. Human participants continue to outperform AI models in some specifically designed tasks, yet recent systems have achieved substantial improvements relative to earlier generations. The competition also showed the growing importance of self-improvement, including test-time adaptation, iterative refinement, and synthetic-data generation, which contribute to performance gains.

Another important metric concerns the length and complexity of tasks that AI systems can successfully complete (Kwa et al. 2026).[8] Scientific research rarely consists of solving isolated problems. Rather, it typically requires sustained reasoning over extended periods, involving multiple steps, and iterative evaluation. Progress in this domain has been particularly evident in software engineering, where AI systems have advanced from completing tasks lasting seconds or minutes to undertaking projects requiring several hours of continuous work. Such developments are highly relevant to scientific applications, which also usually involve extended periods (literature reviews, experimental planning, data analysis and interpretation etc.). This trend has led some authors to propose an analogue of Moore's Law for AI agents (Binksmith et al. 2026)[9], according to which the duration and complexity of tasks that can be completed autonomously continue to increase at an approximately exponential rate.

[7] François Chollet, Mike Knoop, Gregory Kamradt, and Bryan Landers, "ARC Prize 2025: Technical Report," arXiv, 2026, https://doi.org/10.48550/arXiv.2601.10904.
[8] Thomas Kwa, Ben West, Joel Becker, Amy Deng, Katharyn Garcia, Max Hasin, Sami Jawhar, Megan Kinniment, Nate Rush, Sydney von Arx, et al., "Measuring AI Ability to Complete Long Tasks," arXiv, 2025, https://doi.org/10.48550/arXiv.2503.14499.
[9] Adam Binksmith et al., "A New Moore's Law for AI Agents," *AI Digest*, March 24, 2026, accessed June 12, 2026, https://theaidigest.org/time-horizons.

Whether this rate of progress will continue remains uncertain. Even so, the capabilities already achieved have transformed AI from a specialised computational tool into a broadly applicable research technology embedded in scientific practice across a wide range of disciplines.

## 3. The Diffusion of AI Across the Sciences

The rapid growth of AI capabilities has been accompanied by an equally rapid expansion of AI's presence within scientific research. Although AI had already begun spreading beyond computer science prior to the public release of ChatGPT, and the public availability of foundation models in 2022 accelerated this process substantially. AI is now employed across the majority of scientific disciplines, ranging from the natural sciences, engineering, and medicine to the humanities and social sciences. In this respect, AI increasingly resembles earlier general-purpose technologies, such as computers and the internet, whose influence ultimately extended throughout the entire scientific enterprise.

This transformation can be observed through a variety of scientometric indicators. The study *Rise of Generative Artificial Intelligence in Science* (Ding, Lawson, and Shapira 2025)[10] documents this rapid diffusion of generative AI throughout scientific research. Although growth has occurred across disciplines since 2022, it has been particularly pronounced in fields such as the social sciences, psychology, education, and the arts, where the societal implications of AI have themselves become major objects of investigation.

Further comprehensive evidence is provided by Stanford University's *AI Index Report 2026* (Sajadieh et al. 2026).[11] According to the report, AI-related publications in the natural sciences increased by more than one quarter between 2024 and 2025, reaching approximately 80,000 publications in 2025. AI-related research now constitutes a significant and rapidly growing proportion of overall scientific output. Such figures indicate that the incorporation of AI into scientific practice is no longer confined to isolated research communities but has become a widespread phenomenon throughout contemporary science.

Scientists also increasingly use LLMs in scientific writing. Recent analyses based on identifiable LLM-modified text suggest a rapid uptake in AI-assisted writing following 2023 (Liang et al. 2025).[12] Estimates from September 2024 indicate particularly high rates in computer science, where AI-assisted writing may account for as much as 22 per cent of published work. Somewhat lower, though still substantial, estimates have been reported for fields such as statistics, physics, and finally mathematics, where rates may approach 9 per cent. Although such estimates remain methodologically imperfect, they nevertheless provide evidence that AI systems are becoming integrated into everyday scientific practice even in these basic tasks.

---

[10] Liangping Ding, Cornelia Lawson, and Philip Shapira, "Rise of Generative Artificial Intelligence in Science," *Scientometrics* 130 (2025): 5093–5114, https://doi.org/10.1007/s11192-025-05413-z.
[11] Sha Sajadieh, Loredana Fattorini, Raymond Perrault, Yolanda Gil, Vanessa Parli, Lapo Santarlasci, Juan Pava, Nestor Maslej, Russ Altman, and Erik Brynjolfsson, et al., *The AI Index 2026 Annual Report* (Stanford, CA: AI Index Steering Committee, Institute for Human-Centered AI, Stanford University, 2026), https://hai.stanford.edu/assets/files/ai_index_report_2026.pdf.
[12] Weixin Liang, Yuhui Zhang, Zhengxuan Wu, et al., "Quantifying Large Language Model Usage in Scientific Papers," *Nature Human Behaviour* 9 (2025): 2599–2609, https://doi.org/10.1038/s41562-025-02273-8.

Publication data from major scientific publishers corroborate this trend. Across the portfolio of *Nature* journals, the number of papers containing references to artificial intelligence increased more than seventy-fold between 2015 and 2025, surpassing 16,000 publications annually (Nature 2026).[13] The trend is equally visible within the flagship journal *Nature* itself, where occurrences of the term "artificial intelligence" rose from approximately fifty papers per year in 2015 to more than seven hundred in 2025. These figures reflect the continuing application of AI methods as well as the emergence of AI as a central object of scientific inquiry across the majority of disciplines. Whether this development ultimately constitutes a profound transformation of scientific research – or even a paradigmatic shift in the organisation of scientific knowledge production – remains an open question, which will be analysed in the concluding section of the paper.

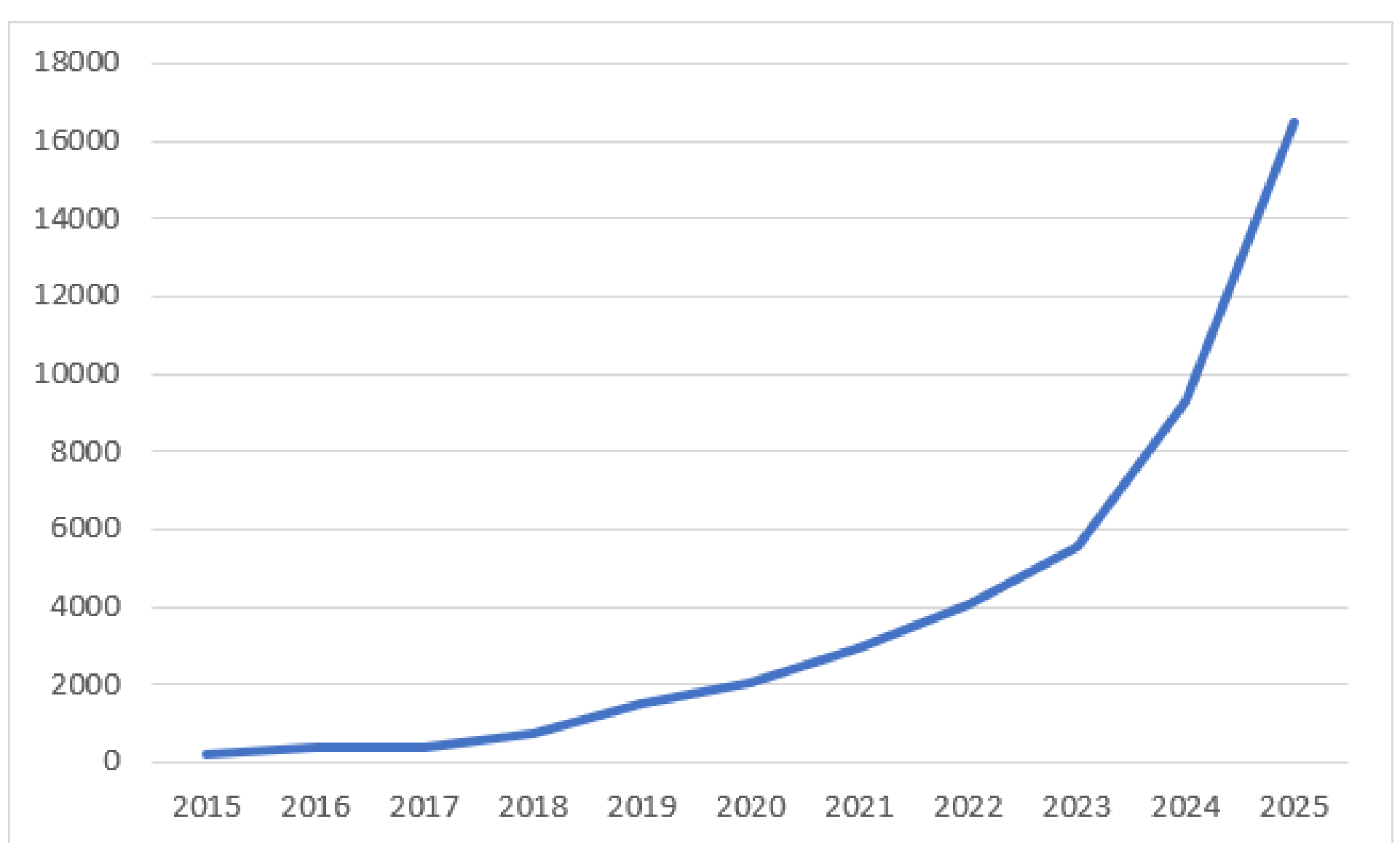

**Chart 1.** Number of articles containing the term "artificial intelligence" in the portfolio of Nature journals.

[13] Nature, "Search results page," accessed June 11, 2026, https://www.nature.com/.

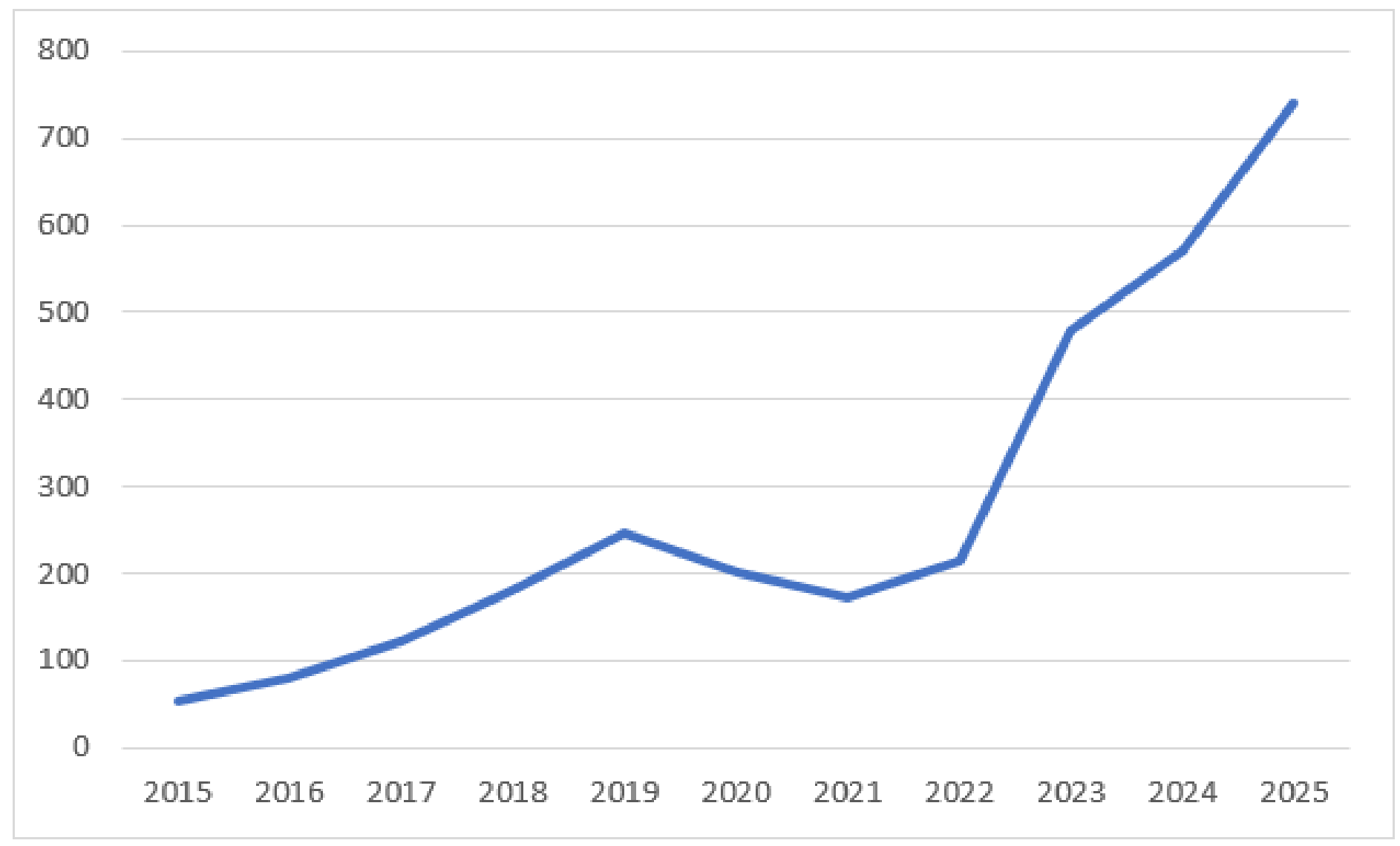

**Chart 2.** Number of articles containing the term "artificial intelligence" in the journal Nature.

## 4. A Typology of AI Systems in Scientific Research

The increasing role of AI in scientific research has been accompanied by the emergence of a heterogeneous pool of computational tools. Before the rise of LLMs, most scientific applications of AI took the form of specialised ML systems designed for narrowly defined tasks. Those systems remain important, but at the same time, foundation models have introduced new forms of participation in research that extend beyond prediction and classification. For analytical purposes, it is therefore useful to distinguish four broad categories of AI systems currently employed in scientific work: Specialized Scientific AI, Scientific AI Assistants, Scientific AI Agents, and Hybrid AI Experimental Systems. The classification proposed here is based primarily on the role AI plays in scientific practice rather than on the underlying algorithms or architectures – the categories are therefore heuristic and not mutually exclusive.

### 4.1. Specialized Scientific AI

The first category consists of specialised systems developed to address specific problems. These are AI systems optimised for narrowly defined scientific tasks, regardless of what algorithm or architecture they use. Many of these tools emerged before the era of publicly available LLMs and remain widely used throughout science. Their capabilities are often highly sophisticated, but they are typically confined to well-defined domains.

A prominent example is AlphaFold. Protein structure determination has traditionally relied on labour-intensive experimental techniques such as X-ray crystallography, NMR spectroscopy, and cryo-electron microscopy, while computational prediction methods achieved only limited success prior to the recent advances enabled by transformer-based architectures. AlphaFold's earliest version, AlphaFold 1, relied primarily on convolutional neural networks (CNN), but later generations, beginning with AlphaFold 2, moved beyond CNN-based approaches and incorporated hybrid architectures with transformer-derived mechanisms and specialised attention modules, including the Evoformer and later the Pairformer, resulting in major improvements in predictive accuracy (Jumper et al. 2021).[14] The evolution of AlphaFold illustrates the broader transition from relatively narrow ML systems toward more universal AI systems.

[14] John Jumper, Richard Evans, Alexander Pritzel, et al., "Highly Accurate Protein Structure Prediction with AlphaFold," *Nature* 596 (2021): 583–589, https://doi.org/10.1038/s41586-021-03819-2.

### 4.2. Scientific AI Assistants

The second category consists of Scientific AI Assistants. These systems are typically based on LLMs and operate primarily through interaction with human users. Unlike traditional specialised AI tools, they are not restricted to a single task but can assist researchers across a broad range of activities.

In the scientific context, AI assistants are commonly used for literature review, information retrieval, hypothesis exploration, data analysis, coding, proofreading, critique, and manuscript preparation. Examples include systems such as ChatGPT Deep Research, Gemini Deep Research, SciSpace, NotebookLM, and Cursor. Although such systems can enhance productivity, they remain fundamentally reactive: they perform tasks in response to requests from human researchers rather than pursuing research goals independently.

At the same time, the boundary between assistants and agents is becoming increasingly porous. Some systems commonly described as assistants – for example ChatGPT Deep Research, Claude Research, Gemini Deep Research, and Perplexity Labs – already assume more agentic functions. In that respect, assistants and agents are best understood not as sharply distinct categories but as points on a continuum of increasing autonomy.

### 4.3. Scientific AI Agents

A third category consists of AI agents developed for scientific applications. In contrast to assistants, agents are capable of carrying out more autonomous and extended tasks. They typically decompose complex objectives into smaller ones, coordinate multiple stages of execution, and may employ several specialised sub-agents working toward a common goal.

From the perspective of scientific discovery, this category is especially significant. Multi-agent systems increasingly perform activities that resemble components of scientific research itself, including literature analysis, hypothesis generation, experiment design, evaluation, critique, and iterative refinement. For this reason, such systems are often described as “AI scientists”, “co-scientists”, “hypothesis machines”, or “virtual scientific collaborators”. Examples include Sakana’s AI Scientist (Lu et al. 2024)[15], Google’s Co-Scientist (Gottweis et al. 2026),[16] and FutureHouse's Robin (Ghareeb et al. 2026).[17] These systems employ teams of interacting agents that collaborate, critique one another, and iteratively improve proposed solutions, while still retaining human expert oversight.

---

[15] Chris Lu, Cong Lu, Robert Tjarko Lange, Jakob Foerster, Jeff Clune, and David Ha, “The AI Scientist: Towards Fully Automated Open-Ended Scientific Discovery,” arXiv, 2024, https://doi.org/10.48550/arXiv.2408.06292.

Their objective is to assist human researchers but also to automate substantial portions of the scientific process. Although current systems continue to operate under human supervision and can be tasked in natural language, they nevertheless represent an important step toward increasingly autonomous forms of scientific research.

### 4.4. Hybrid AI Experimental Systems

A fourth category consists of Hybrid AI Experimental Systems. These could be treated as a subcategory of agentic systems, but because they extend beyond pure computation into the physical world, they merit separate category. Such systems combine AI-driven reasoning with robotic platforms capable of conducting experiments and manipulating laboratory equipment, thereby functioning as autonomous or "self-driving" laboratories.

The emergence of these systems marks an important development in the history of scientific automation. Earlier forms of AI primarily assisted with information processing and analysis. Hybrid computational-physical systems, by contrast, increasingly participate directly in experimental research itself. Thus, they provide a glimpse of a future in which substantial portions of scientific discovery may be conducted through such integrated platforms.

One example is ChemAgents, a robotic AI chemist based on a hierarchical multi-agent system incorporating the Llama-3.1-70B model (Song et al. 2025).[18] This autonomous platform integrates LLMs, planning modules, automated instrumentation, and robotic execution. It is capable of reviewing scientific literature, designing experiments, operating laboratory equipment, collecting data, analysing results, and refining subsequent experimental strategies.

Similarly, the Recursion platform exemplifies a "self-driving" wet-and-dry biochemical laboratory. It combines robotic operations, digital twins, natural-language-driven lab control, and uses large biological and chemical datasets, and SOTA multimodal foundation models

---

[16] Juraj Gottweis, Wei-Hung Weng, Alexander Daryin et al., "Accelerating Scientific Discovery with Co-Scientist," *Nature* (2026), https://doi.org/10.1038/s41586-026-10644-y.
[17] Ali Essam Ghareeb, Benjamin Chang, Ludovico Mitchener, et al., "A Multi-Agent System for Automating Scientific Discovery," *Nature* (2026), https://doi.org/10.1038/s41586-026-10652-y.
[18] Tao Song, Man Luo, Xiaolong Zhang, LinJiang Chen, Yan Huang, Jiaqi Cao, Qing Zhu, Daobin Liu, Baicheng Zhang, Gang Zou, et al., "A Multiagent-Driven Robotic AI Chemist Enabling Autonomous Chemical Research On Demand," *Journal of the American Chemical Society* 147, no. 15 (2025): 12534–12545, https://doi.org/10.1021/jacs.4c17738.

running on BioHive-2 (a platform built on NVIDIA chips). Recursion performs millions of multi-omic experiments each week and aims to increase efficiency across the biopharmaceutical pipeline through *in silico* modelling, target identification, automated experimentation, and analysis. It also relies on large biological and chemical datasets, including perturbational data (Wenkel et al. 2026)[19]

The typology outlined above reflects increasing degrees of autonomy and scientific participation. Specialised systems still remain important, but contemporary AI now extends well beyond narrow tasks toward hypothesis generation, experimental design, and partially autonomous research.

[19] Frederik Wenkel, Wilson Tu, Cassandra Masschelein, et al., "TxPert: Using Multiple Knowledge Graphs for Prediction of Transcriptomic Perturbation Effects," *Nature Biotechnology* (2026), https://doi.org/10.1038/s41587-026-03113-4.

## 5. AI and Scientific Discovery Across Disciplines

The examples discussed in the following sections illustrate how different categories of AI systems are already contributing to scientific progress across a range of disciplines. For present purposes, this section is organised into four broad areas: mathematics and computer science; physics, chemistry, and materials science; life sciences; and the behavioural and social sciences. It focuses on a number of salient and concrete contributions, such as proved or disproved theorems, new algorithms, experimentally validated hypotheses, and therapeutic candidates in later stages of development. Its purpose is not to provide a comprehensive account of each discipline, but to indicate the present range of AI capabilities.

### 5.1. Mathematics and Computer Science

Among all scientific disciplines, mathematics and computer science provide some of the clearest evidence of the rapid improvement of AI capabilities. Only a few years ago, many experts regarded LLMs as unsuitable for advanced mathematical work and incapable of producing results comparable to those of professional mathematicians. Recent developments have altered that assessment. Frontier models are increasingly able not only to solve difficult problems but also to assist in original research, proof generation, and the discovery of previously unknown results.

One important illustration is AlphaProof Nexus (Tsoukalas et al. 2026),[20] a Google DeepMind framework that combines Gemini-based prover agents with formal verification tools such as Lean. The system autonomously solved several open Erdős problems and demonstrated the usefulness of AI for research-level mathematics. Its most important feature is the integration of formal verification, which allows proposed proofs to be checked rigorously and thereby mitigates one of the traditional weaknesses of LLMs. The framework can also be applied across areas such as combinatorics, optimisation, graph theory, algebraic geometry, and quantum optics, showing that such systems are not limited to a single subdiscipline.

Recent work also shows a shift toward genuine research problems. One example is the resolution of a long-standing problem in combinatorial geometry originally posed by Erdős. An OpenAI general-purpose reasoning model, rather than a narrowly specialised mathematical system, produced a solution that was subsequently examined by professional mathematicians and found to involve unexpected connections between elementary geometry and algebraic number theory (OpenAI 2026).[21] Timothy Gowers described the result as a milestone in AI-assisted mathematics. Related systems now combine informal reasoning agents with formal proof assistants; for instance, Ju et al. (2026)[22] report the autonomous resolution and Lean 4 verification of an open problem in commutative algebra.

[20] George Tsoukalas, Anton Kovsharov, Sergey Shirobokov, et al., "Advancing Mathematics Research with AI-Driven Formal Proof Search," arXiv, 2026, https://doi.org/10.48550/arXiv.2605.22763.

A similar pattern appears in computer science. AI systems have begun contributing to theoretical research and algorithmic discovery. Anthropic's Claude Opus assisted in resolving a graph-theoretic conjecture concerning Hamiltonian cycles, which Knuth described as evidence of a dramatic advance in automated deduction and creative problem solving (Knuth 2026).[23] Google's AlphaEvolve, an evolutionary coding agent built on Gemini, has generated verifiably improved algorithms, including a more efficient procedure related to complex-valued matrix multiplication and Strassen's algorithm (Novikov et al. 2025).[24] FunSearch similarly combined an LLM with a systematic evaluator capable of filtering errors and fabrications, and used this architecture to generate new solutions to the cap-set problem in extremal combinatorics (Romera-Paredes, Barekatain, & Novikov 2024).[25] Although these methods rely heavily on computational exploration, they show that AI systems can produce novel mathematical and algorithmic structures rather than only reproduce existing knowledge.

Other systems aim to automate larger parts of the research process itself. Sakana's AI Scientist can generate ideas, write code, conduct computational experiments, analyse results, and produce complete manuscripts. In one evaluation, system-generated manuscripts passed peer review at a machine-learning workshop, which the authors described as an "AI Scientist Turing Test" (Lu et al. 2026) [26].Taken together, these cases suggest that current AI systems, despite continuing dependence on human supervision, can now contribute to mathematical discovery at a level that would have seemed implausible only recently.

---

[21] OpenAI, "OpenAI Model Disproves Discrete Geometry Conjecture," accessed June 11, 2026, https://openai.com/cs-CZ/index/model-disproves-discrete-geometry-conjecture/. (It is important to note that these solutions of Erdős open problems have not been yet published in a peer-reviewed journal.)
[22] Haocheng Ju, Guoxiong Gao, Jiedong Jiang, Bin Wu, Zeming Sun, Leheng Chen, Yutong Wang, Yuefeng Wang, Zichen Wang, Wanyi He, et al., "Automated Conjecture Resolution with Formal Verification," arXiv, 2026, https://doi.org/10.48550/arXiv.2604.03789.
[23] Donald E. Knuth, "Claude Cycles," PDF manuscript, accessed June 11, 2026, https://www-cs-faculty.stanford.edu/~knuth/papers/claude-cycles.pdf.
[24] Alexander Novikov, Ngân Vũ, Marvin Eisenberger, Emilien Dupont, Po-Sen Huang, Adam Zsolt Wagner, Sergey Shirobokov, Borislav Kozlovskii, Francisco J. R. Ruiz, Abbas Mehrabian, et al., "AlphaEvolve: A Coding Agent for Scientific and Algorithmic Discovery," arXiv, 2025, https://doi.org/10.48550/arXiv.2506.13131.
[25] Bernardino Romera-Paredes, Mohammadamin Barekatain, Alexander Novikov, et al., "Mathematical Discoveries from Program Search with Large Language Models," *Nature* 625 (2024): 468–475, https://doi.org/10.1038/s41586-023-06924-6.
[26] Chris Lu, Cong Lu, Robert Tjarko Lange, et al., "Towards End-to-End Automation of AI Research," *Nature* 651 (2026): 914–919, https://doi.org/10.1038/s41586-026-10265-5.

### 5.2. Physics, Chemistry, and Materials Science

Similar developments are visible in the physical sciences. In physics, the most important trend is not the solution of isolated problems as by the previous domain-specific ML applications but the emergence of more general AI systems capable of exploring, and modelling complex physical phenomena. In this respect, the physical sciences are beginning to develop scientific foundation models analogous to those that transformed natural-language processing.

One example comes from theoretical physics, where frontier language models have assisted with sophisticated mathematical reasoning. GPT-5.2 Pro helped simplify expressions describing gluon interactions and identify a generalised mathematical formulation later verified by researchers. An OpenAI model subsequently generated a proof whose validity was confirmed analytically (Guevara et al. 2026).[27] These examples do not yet amount to autonomous discovery in a strong sense, because researchers still verify and interpret the outputs. They nevertheless show that contemporary AI systems can participate in advanced theoretical reasoning in fields traditionally regarded as highly specialised.

A broader change is represented by foundation models for physics. Traditional ML systems in physics were usually trained for narrow problem classes and often required retraining when applied to new systems, boundary conditions, or governing equations. The General Physics Transformer, GPhyT (Wiesner et al. 2026),[28] seeks to overcome this limitation by training on a large and diverse collection of simulation data. It can model phenomena such as fluid dynamics, shock waves, and thermal convection, maintain stable long-horizon predictions, and generalise to previously unseen physical conditions. These results point toward general-purpose modelling systems for the physical sciences, or what the authors describe as a possible "universal physics engine".

In chemistry, AI systems are moving beyond prediction toward experimental participation. ChemAgents combines an LLM with a robotic platform in an integrated system able to review literature, design experiments, perform robotic laboratory procedures, analyse results, and refine subsequent experiments (Song et al. 2025).[29] It was used to discover metal-organic high-entropy catalysts comprising five metallic elements, conducting syntheses and measurements, analysing data from 100 automated experiments, and optimising the most suitable composition. The system therefore demonstrate how multi-agent robotic chemists may support on-demand autonomous chemical research with limited human intervention.

[27] Alfredo Guevara, Alexandru Lupsasca, David Skinner, Andrew Strominger, and Kevin Weil, "Single-Minus Gluon Tree Amplitudes Are Nonzero," arXiv, 2026, https://doi.org/10.48550/arXiv.2602.12176.
[28] Florian Wiesner, Zoë J. Gray, Matthias Wessling, and Stephen Baek, "Towards a Physics Foundation Model," arXiv, 2025, https://arxiv.org/abs/2509.13805.
[29] Song, "A Multiagent-Driven Robotic AI Chemist Enabling Autonomous Chemical Research On Demand".

Foundation models are also emerging in chemistry and materials science. SMI-TED, trained on datasets comprising 91 million molecular sequences, supports tasks such as molecular-property prediction and reaction-outcome estimation (Soares et al. 2025).[30] In materials science, atomistic models such as MACE-MP use *ab initio* methods, including density functional theory, to predict atomic interactions across solids, liquids, gases, interfaces, and molecular systems (Batatia et al. 2025).[31] These systems represent a shift away from narrowly optimised tools toward more general scientific collaborators capable of modelling, simulations, and experimental design across several exact-sciences.

### 5.3. Life Sciences

The life sciences offer some of the most compelling examples of AI's transformative potential. Biological systems are complex, and organised on multiple interacting levels, which makes research especially laborious and time-consuming. AI has therefore become valuable both for solving specific bottlenecks and, increasingly, for participating in several stages of discovery, from prediction and interpretation to hypothesis generation and experimental planning.

This shift is epitomised by protein structure prediction. Because protein function depends critically on three-dimensional structure, determining protein configurations is central to understanding cellular processes, disease mechanisms, and drug interactions. AlphaFold transformed this field by improving structure prediction directly from amino-acid sequences and reducing work that previously required months or years to a matter of minutes. AlphaFold 3 expanded the scope beyond individual protein structures to interactions among proteins, DNA, RNA, and smaller biological molecules, improving modelling of antibody-antigen complexes, protein-nucleic-acid binding, and small-molecule drug binding (Abramson et al.

---

[30] Eduardo Soares, Emilio Vital Brazil, Victor Shirasuna, et al., "An Open-Source Family of Large Encoder-Decoder Foundation Models for Chemistry," *Communications Chemistry* 8 (2025): 193, https://doi.org/10.1038/s42004-025-01585-0.

[31] Ilyes Batatia, Philipp Benner, Yuan Chiang, Alin M. Elena, Dávid P. Kovács, Janosh Riebesell, Xavier R. Advincula, Mark Asta, Matthew Avaylon, William J. Baldwin, et al., "A Foundation Model for Atomistic Materials Chemistry," *The Journal of Chemical Physics* 163, no. 18 (2025): 184110, https://doi.org/10.1063/5.0297006.

2024).[32] Earlier versions of AlphaFold have already had tangible scientific value. In malaria research, a combination of experimental work and AlphaFold 2 prediction made it possible to obtain the molecular structure of Pfs48/45, a gamete-surface protein essential to the parasite's development in the mosquito (Ko et al. 2022).[33] This target had resisted leading experimental methods, including X-ray crystallography and cryo-electron microscopy. The resulting structure can inform vaccine design by indicating which parts of the protein to use and how they might be organised.

Other systems extend AI's role from prediction to active discovery. Google's Co-Scientist, a Gemini-based multi-agent system, can analyse literature, synthesise knowledge, generate hypotheses, propose protocols, and refine outputs through internal critique and evaluation. Importantly, it remains within a scientist-in-the-loop workflow, so researchers continue to guide and assess the discovery process. In biomedical applications, it generated hypotheses concerning drug repurposing for acute myeloid leukemia and identified candidates that showed *in vitro* efficacy. It also proposed a mechanism of mobile genetic element transfer between bacteria relevant to antimicrobial resistance, later independently confirmed by a human-led project (Gottweis et al. 2026).[34]

Robin represents a similar approach. This multi-agent platform combines literature analysis, data interpretation, hypothesis generation, experimental planning, and repeated evaluation to identify testable therapeutic strategies. Its authors report substantial time savings; for example, the analysis of hundreds of papers took about 30 minutes rather than an estimated 294 hours of human labour. In one demonstration involving dry age-related macular degeneration, Robin identified the repurposed candidate Ripasudil and the novel candidate KL001, both of which produced encouraging *in vitro* results (Ghareeb et al. 2026).[35]

The next stage involves integrating AI into experimental infrastructure. For instance, the Recursion platform combines large biological datasets, robotic experimentation, and foundation models to accelerate target identification, compound screening, and drug development. Through collaborations, it has advanced several therapeutic candidates into clinical testing. One case concerns familial adenomatous polyposis, a rare genetic disorder involving precancerous gastrointestinal polyps (Samadder 2026).[36] For FAP patients, who face high colorectal-cancer risk and have no approved therapies beyond colectomy, the platform helped identify disease-relevant morphological features, possible therapeutic mechanisms, and candidate interventions now in phase 1B and 2 of clinical trials.

[32] Josh Abramson, Jonas Adler, Jack Dunger, et al., "Accurate Structure Prediction of Biomolecular Interactions with AlphaFold 3," *Nature* 630 (2024): 493–500, https://doi.org/10.1038/s41586-024-07487-w.
[33] Kuang-Ting Ko, Frank Lennartz, David Mekhaiel, Bora Guloglu, Arianna Marini, Danielle J. Deuker, Carole A. Long et al., "Structure of the Malaria Vaccine Candidate Pfs48/45 and Its Recognition by Transmission Blocking Antibodies," *Nature Communications* 13, no. 1 (2022): 5603, https://doi.org/10.1038/s41467-022-33379-6.
[34] Juraj Gottweis "Accelerating Scientific Discovery with Co-Scientist".
[35] Ali Essam Ghareeb, "A Multi-Agent System for Automating Scientific Discovery".

These examples show that AI is already contributing to life-science research beyond data processing and routine automation. It can accelerate discovery by generating hypotheses, supporting complex reasoning, analysing results, and in some cases by performing parts of experimental work itself. These achievements, however, must be considered alongside the limitations discussed below.

### 5.4. Behavioural and Social Sciences

The behavioural and social sciences have also adopted AI both as an object of inquiry and as a research instrument. In these fields, AI systems are used to investigate cognition, personality, social judgement, and persuasion. Recent work includes foundation-style models of cognition as well as studies of AI-mediated persuasion. In the latter cases the AI is used not only to model and predict human behaviour, as in the physical or chemical sciences, but also to influence human judgement and behaviour.

One important line of research concerns general models of cognition. Binz et al. (2025)[37] developed Centaur, a model intended to predict human behaviour across tasks expressible in natural language. It was created by fine-tuning a state-of-the-art Llama model on data from more than 60,000 participants across 160 experiments. Centaur outperformed existing domain-specific cognitive models and generalised to unseen or modified tasks in logical reasoning, moral decision-making, and economic games. It predicts not only average behaviour but distributions over population-level trajectories, pointing toward forms of "automated cognitive science" in which experiments can be explored in silico before or alongside human testing.

[36] Jewel Samadder, "Ongoing Phase 1b/2 Trial of the Allosteric MEK1/2 Inhibitor REC-4881 as Monotherapy in Familial Adenomatous Polyposis (FAP): Preliminary Safety and Efficacy Data," presentation, May 4, 2025, Recursion Pharmaceuticals, PDF, https://ir.recursion.com/static-files/cef86a12-64aa-4ff9-9074-2f81fb3efd17.
[37] Marcel Binz, Elif Akata, Matthias Bethge, et al., "A Foundation Model to Predict and Capture Human Cognition," *Nature* 644 (2025): 1002–1009, https://doi.org/10.1038/s41586-025-09215-4.

Other work examines whether LLMs can represent human agents and track mental states. Strachan et al. (2024)[38] found that models often displayed human-comparable performance on theory-of-mind tasks involving indirect requests, false beliefs, and deliberate misdirection. A related psychometric study showed that AI systems can predict correlations between personality-questionnaire items more accurately than laypeople and most individual academic experts. In that study, frontier LLMs outperformed most individual human participants, while PersonalityMap, a specialised system trained on pairs of psychometric items, matched expert group-level performance on more domain-specific tasks (Schoenegger et al. 2025).[39]

A second line of research shifts from prediction to influence. Costello et al. (2024)[40] found that conversational generative AI could significantly reduce conspiratorial beliefs, with effects persisting for months. This suggests beneficial applications in public communication and education, but it also illustrates AI's capacity to shape beliefs and attitudes through sustained, personalised interaction. Other studies point to risks. Glickman and Sharot (2025)[41] reported that repeated interaction with biased AI systems can alter perceptual, emotional, and social judgements in ways that strengthen human bias. Salvi et al. (2025)[42] similarly found that LLMs can generate persuasive content, especially with microtargeting where outputs are conditioned on personal attributes and psychological profiles; in such settings, GPT-4 outperformed human opponents when given access to participants' sociodemographic data.

Overall, work in the behavioural and social sciences suggests that current AI systems can analyse and predict human behaviour, but also exploit what they learn about belief formation and judgement in ways that enable persuasion and, potentially, manipulation. These disciplines are therefore important not only for assessing the epistemic promise of AI in science but also for understanding its ethical and political risks.

Across the disciplines surveyed above, common patterns emerge. In the formal and exact sciences, AI contributes most when tightly coupled to the structure of the problem. In the life sciences, its impact is greatest where systems integrate long-context information, multimodal representations, and experimental feedback, functioning not only as predictive tools but also as instruments of functional interpretation and design.

---

[38] James W. A. Strachan, Dalila Albergo, Giulia Borghini, et al., "Testing Theory of Mind in Large Language Models and Humans," *Nature Human Behaviour* 8 (2024): 1285–1295, https://doi.org/10.1038/s41562-024-01882-z.

[39] Philipp Schoenegger, Spencer Greenberg, Alexander Grishin, et al., "AI Can Outperform Humans in Predicting Correlations Between Personality Items," *Communications Psychology* 3 (2025): 23, https://doi.org/10.1038/s44271-025-00205-w.

[40] Thomas H. Costello, Gordon Pennycook, and David G. Rand, "Durably Reducing Conspiracy Beliefs through Dialogues with AI," *Science* 385, no. 6714 (2024): eadq1814, https://doi.org/10.1126/science.adq1814.

[41] Moshe Glickman and Tali Sharot, "How Human–AI Feedback Loops Alter Human Perceptual, Emotional and Social Judgements," *Nature Human Behaviour* 9 (2025): 345–359, https://doi.org/10.1038/s41562-024-02077-2.

[42] Francesco Salvi, Manoel Horta Ribeiro, Riccardo Gallotti, et al., "On the Conversational Persuasiveness of GPT-4," *Nature Human Behaviour* 9 (2025): 1645–1653, https://doi.org/10.1038/s41562-025-02194-6 .

## 6. Current Limitations of AI in Scientific Discovery

Despite the progress of machine learning and large language models, their application to scientific discovery remains constrained by a number of limitations. Some arise from the current capabilities of the models themselves, whereas others stem from the nature of scientific inquiry and from the institutional conditions under which research is conducted.

### 6.1. General Limitations of Contemporary AI Systems

Among the most persistent technical shortcomings are hallucinations, factual inaccuracies, and difficulties in reliably distinguishing trustworthy information from false or fabricated content. Although these problems have been reduced in recent generations of models, they have by no means disappeared. A striking example is the case of the apparently fabricated disease "Bixonimania", which appeared in preprints attributed to a non-existent researcher and was subsequently propagated not only by AI systems but also by human researchers relying on AI-generated outputs (Stokel-Walker 2026).[43] Such cases undermine the reliability of otherwise capable systems, which can contribute to the dissemination of unreliable scientific content when their outputs are not subjected to adequate scrutiny.

A further challenge concerns the evaluation of scientific capabilities. Contemporary models often perform impressively on established benchmarks, yet benchmark success does not necessarily translate into effective participation in genuine scientific research. Song et al. (2026)[44] argue that strong performance on decontextualised tasks can obscure weaknesses that become apparent when models are required to formulate hypotheses, design experiments or simulations, and interpret results within broader research programmes. To investigate these issues, they developed the Scientific Discovery Evaluation framework SDE), which assesses end-to-end scientific-research tasks across biology, chemistry, materials science, and physics. Within this setting, state-of-the-art models exhibited reasoning limitations, performance plateaus, diminishing returns from scaling, and a variety of systematic weaknesses. At the same time, the experimental conditions themselves were not necessarily optimal, suggesting that such findings should be interpreted cautiously rather than as definitive limits.

---

[43] Chris Stokel-Walker, "Scientists Invented a Fake Disease. AI Told People It Was Real," *Nature* 652, no. 8110 (2026): 559–561, https://doi.org/10.1038/d41586-026-01100-y.
[44] Zhangde Song, Jieyu Lu, Yuanqi Du, Botao Yu, Thomas M. Pruyn, Yue Huang, Kehan Guo, Xiuzhe Luo, Yuanhao Qu, Yi Qu, et al. "Evaluating Large Language Models in Scientific Discovery." arXiv preprint arXiv:2512.15567. Version 2, May 8, 2026. https://doi.org/10.48550/arXiv.2512.15567.

Current systems also remain constrained by deficiencies in memory and learning. Persistent memory, long-horizon reasoning, and the maintenance of coherent performance over time continue to present major challenges. More fundamentally, contemporary models are generally unable to engage in robust forms of continual learning and self-improvement after deployment. As Hassabis (Kantrowitz 2026)[45] has noted, advances in continual learning, larger context windows, selective memory mechanisms, and more reliable forms of long-term memory remain important goals for current AI research. Related limitations include insufficient robustness across environments, restricted generalisability, limited autonomous behaviour, and persistent difficulty in integrating AI systems into complex real-world research.

There are also more general constraints on future progress. One widely discussed concern is the emergence of a "data wall", namely the possibility that the supply of high-quality training data may become insufficient to sustain current rates of improvement. Synthetic data may partially alleviate this problem, but it is unlikely to eliminate it entirely. Other potential constraints include computational requirements, memory limitations, and shortcomings of current ML paradigms and transformer architectures themselves. The persistence of errors even in relatively simple tasks has also led some researchers to argue that present systems still lack a sufficiently rich understanding of the physical world. Dawid and LeCun (2023),[46] for example, interpret such failures as evidence of the absence of a genuine world model, insufficient physical grounding[47], and limited long-term planning abilities. Progress may also be constrained by external factors, including regulation and voluntary restrictions adopted by AI companies in response to safety and alignment concerns.

At the same time, not all criticisms remain equally compelling in light of recent developments. In particular, increasingly strong versions of the view that AI systems merely reproduce existing knowledge without generating genuinely novel outputs are now harder to maintain in their original form. The best-known formulation of this critique is the "stochastic parrot" thesis, according to which language models do not possess genuine understanding but merely recombine linguistic patterns encountered during training (Bender et al. 2021).[48] The mathematical and scientific examples discussed earlier do not settle the broader philosophical debate about understanding or originality, but they do refute the strongest versions of the claim that state-of-the-art systems function only as mechanisms of recombination. At a minimum, they suggest that such systems can now produce outputs that satisfy ordinary standards of novelty.

---

[45] Alex Kantrowitz, "Google DeepMind CEO Demis Hassabis: AI's Next Breakthroughs, AGI Timeline, Google's AI Glasses Bet," *Big Technology Podcast*, YouTube video, premiered January 23, 2026, https://www.youtube.com/watch?v=bgBfobN2A7A.

[46] Anna Dawid and Yann LeCun, "Introduction to Latent Variable Energy-Based Models: A Path Towards Autonomous Machine Intelligence," arXiv, 2023, https://doi.org/10.48550/arXiv.2306.02572.

[47] However, this argument can be successfully contested, cf. Blaise Agüera y Arcas, "Do Large Language Models Understand Us?" Daedalus 151, no. 2 (2022): 183–197, https://doi.org/10.1162/DAED_a_01909.

### 6.2. Limitations of Current Scientific AI Systems

Beyond these general limitations, a number of constraints are specific to the actual use of current AI systems in scientific settings.

In the case of AlphaFold, for example, one of its creators, Jumper (Science Friday 2026),[49] has emphasised that the field remains "early in the story". Although AlphaFold has transformed structural biology and accelerated protein-structure prediction, it addresses only one component of the much larger drug-discovery process. Drug development remains constrained by problems of toxicity, solubility, experimental validation, clinical testing, regulatory approval, and numerous other bottlenecks. Consequently, even major advances in protein prediction cannot by themselves eliminate the broader difficulties inherent in biomedical research. As Jumper has cautioned, AlphaFold does not mean that biology has been "solved" or that drug design has become straightforward. Rather, it accelerates specific stages within a much larger and more complex scientific pipeline. By his estimate[50], the technology has made structural biology approximately 5–10 per cent more efficient.

Google's Co-Scientist illustrates a different set of limitations. A significant concern is its dependence on open-access scientific literature. As a result, potentially important hypotheses, theories, datasets, and findings contained in restricted or proprietary sources may remain inaccessible to the system. Moreover, because the scientific literature itself contains irreproducible results and methodological shortcomings, such weaknesses may be inherited by AI systems trained on, or heavily reliant upon, those resources as the Co-Scientist's creators themselves note (Gottweis et al. 2026).[51] This dependence may help explain why Co-

---

[48] Emily M. Bender, Timnit Gebru, Angelina McMillan-Major, and Shmargaret Shmitchell, "On the Dangers of Stochastic Parrots: Can Language Models Be Too Big?," in *Proceedings of the 2021 ACM Conference on Fairness, Accountability, and Transparency* (New York: Association for Computing Machinery, 2021), 610–623, https://doi.org/10.1145/3442188.3445922.
[49] Science Friday. "How AlphaFold Has Changed Biology Research, 5 Years On." YouTube video. Accessed June 11, 2026. https://www.youtube.com/watch?v=n1rr36vHG8A.
[50] This estimate is not yet based on any systematic metrics.
[51] Gottweis, "Accelerating Scientific Discovery with Co-Scientist".

Scientist's results have also been criticised on grounds of limited novelty. In the case of drug repurposing, Lowe (2026)[52] argues that the role of human input, judgement, and oversight was substantially greater than the Co-Scientist paper acknowledges, beginning with the framing of the problem and the selection of candidate solutions. More importantly, Lowe maintains that the targeting of a specific pathway in acute myeloid leukaemia, which Co-Scientist presented as a novel suggestion, was in fact not new, since the line of research had already been investigated previously and was not cited by the authors. Similar reservations apply to some of the other putatively de novo results described in the paper.

Comparable problems have been identified in other AI research assistants, including Sakana's AI Scientist. Evaluations have shown that such systems can generate underdeveloped research ideas, implementation errors, inaccurate citations, and other forms of hallucination (Lu et al. 2026).[53] Those assessments nevertheless concluded that AI Scientist represented an important step toward the automation of scientific discovery and that it was capable of producing research manuscripts with minimal human intervention. At the same time, they revealed substantial weaknesses. AI Scientist frequently misidentified generated ideas as novel, lacked robustness in experiment execution, and produced a considerable number of failed or flawed experiments owing to coding errors. More broadly, it struggled with methodological soundness and demonstrated only a limited capacity to critically evaluate its own outputs. These findings suggest that, despite impressive advances, current AI systems remain far from replacing human scientific judgement and continue to require extensive supervision.

Such limitations indicate that current AI systems function less as standalone autonomous discoverers as they continue to require substantial guidance and remain dependent not only on human scientists for problem formulation and evaluation, but also on the quality of the surrounding scientific literature, institutions, and research infrastructure. Similar limitations affect the deployment of AI systems in other scientific domains as well.

[52] Derek Lowe, "Evaluating 'Co-Scientist', a New AI Science System," *In the Pipeline* (blog), *Science*, May 27, 2026, https://www.science.org/content/blog-post/evaluating-co-scientist-new-ai-science-system.
[53] Lu et al., "Towards End-to-End Automation of AI Research".

## 7. Risks of AI-Driven Scientific Discovery

The growing integration of AI into scientific research introduces not only new opportunities but also new risks. Science and technology have always been dual-use, associated both with major advances and with significant destructive potential, and the integration of AI is likely to intensify this ambivalence.

The relationship between knowledge and power has historically been precarious (Bacon 1597).[54] So the question surrounding the deployment of current or future AI is what consequences may follow if scientific knowledge is generated by systems whose goals, reasoning processes, or decision procedures are only partially understood and which are potentially unaligned.

These risks differ substantially in their nature and severity. Broadly speaking, they can be divided into two categories. The first consists of relatively ordinary and potentially manageable risks associated with the use of current systems. The second encompasses more serious long-term risks that may arise from increasingly autonomous and capable AI systems, especially in scenarios involving advanced agency, and self-improvement.

### 7.1. Ordinary and Mitigable Risks

#### 7.1.1. Epistemic Risks

Many of the immediate risks associated with AI in science arise from its epistemic weaknesses and from the widespread use of AI-generated content by scientists themselves. These include the production of inaccurate information, low-quality scientific content, and fabricated citations (Topaz et al. 2026).[55] One example is the previously discussed case of "Bixonimania", an apparently fabricated disease that originated in preprints attributed to a non-existent researcher and was subsequently reproduced not only by several AI systems (including Copilot, Gemini, Perplexity, and ChatGPT) but also by human researchers relying on AI-generated outputs (Stokel-Walker 2026).[56]

---

[54] Francis Bacon, Essayes. Religious Meditations. Places of Perswasion and Disswasion. Seene and Allowed (London: Printed by John Windet for Humfrey Hooper, 1597), accessed June 12, 2026, Internet Archive, https://archive.org/details/bim_early-english-books-1475-1640_essayes-religious-medit_bacon-francis-viscount_1597.

[55] Maxim Topaz, Nir Roguin, Palavi Gupta, et al., "Fabricated Citations: An Audit across 2.5 Million Biomedical Papers," The Lancet 407 (2026): 1779–1781.

[56] Stokel-Walker, "Scientists Invented a Fake Disease. AI Told People It Was Real".

Other concerns have already been voiced explicitly within scientific communities. The Leiden Declaration on Artificial Intelligence and Mathematics (Alper et al. 2026)[57], for example, warns among other things against the risks that AI poses to correctness, rigour, and standards of proof. Although such problems are not unique to AI, the scale and speed with which AI systems can produce and disseminate information may greatly amplify their impact.

Bias represents another major concern. AI systems inevitably inherit many of the biases present in their training data, including linguistic, cultural, and social biases. These may manifest in forms such as stereotyping, unequal representation, or systematic distortions of scientific knowledge. In addition, new biases may arise through architectural decisions, training procedures, or deliberate interventions during model development. The increasing use of AI-generated scientific content therefore raises important questions about quality control, transparency, and epistemic reliability.

#### 7.1.2. Risks to Science as a Social Institution

Beyond immediate epistemic concerns, some of the most important risks associated with AI in science arise from its interaction with existing scientific practices and institutions. One such risk concerns the disruption of the scientific training pipeline. If increasingly capable AI systems begin to replace work traditionally performed by doctoral students, postdoctoral researchers, or other early-career scientists, the long-established mechanisms through which scientific expertise is transmitted may be weakened.

Scientific training involves much more than the acquisition of technical knowledge. It also includes the transmission of tacit knowledge, methodological judgement, professional norms, and shared scientific values. Much of scientific practice depends upon forms of knowledge that cannot be fully formalised or codified. The replacement of junior researchers by AI systems could therefore undermine the processes through which new generations of scientists acquire the skills necessary for independent research. Similar developments have already appeared in other professions, including law and software development, where automation has reduced demand for certain entry-level roles. A comparable trend in science could produce a paradoxical situation in which scientific productivity increases while opportunities for human participation, training, and career development decline.

---

[57] J. Alper, M. Barany, A. Chavarri Villarello, S. Dahmen, W. Dean, K. Ganapathy, M. Harris, D. Holmes, M. Jamnik, S. Kelk, B. Kra, U. Martin, B. Naskręcki, R. Ochigame, J. Portegies, and J. Schmitt, *Leiden Declaration on Artificial Intelligence and Mathematics* (Zenodo, 2026), https://doi.org/10.5281/zenodo.20302944.

At present, the most vulnerable positions are probably early-career roles. With more capable and more autonomous AI systems, however, including systems acting in the physical world, the range of scientific positions put at risk could become considerably broader, extending across disciplines and across the scientific hierarchy, potentially affecting senior scientists as well. On the other hand, if AI systems are reasonably integrated into existing scientific institutions and processes – provided they remain regulated and designed to support human researchers rather than being evaluated solely in terms of effectiveness – they could facilitate not only an increase in overall scientific output but also growth in scientific employment. For this reason, maintaining meaningful human involvement, especially during the early stages of scientific careers, and carefully integrating AI tools into the research process become important institutional challenges and long-term goals.

### 7.1.3. Security Risks Associated with AI Systems

The increasing capabilities of advanced AI systems have also generated concerns regarding cybersecurity and biosecurity risks.

Cybersecurity concerns have been raised in relation to recent frontier models developed by leading AI companies. Highly capable systems such as Claude Mythos Preview or OpenAI's GPT5.5 can be deployed as autonomous agents capable of identifying vulnerabilities, planning multi-step actions, and interacting with digital environments. While such capabilities may be used defensively, they may also be exploited for malicious purposes (Anthropic 2026).[58] Recent assessments suggest that advanced systems may substantially increase the effectiveness of cyberattacks against poorly defended infrastructures. According to Anthropic, models capable of autonomous operation may pose significant risks. Such risks arise both from deliberate misuse by human actors and from unintended behaviour in increasingly autonomous systems. Related concerns have been identified by the AI Security Institute (2026),[59] whose evaluations showed that advanced models can successfully perform complex attacks involving reconnaissance, exploitation of vulnerabilities, privilege escalation, and ultimately extensive control over simulated corporate networks. These developments have already prompted discussions in the United States about stricter oversight of highly capable systems, particularly where critical infrastructure may be involved (Uren 2026).[60]

---

[58] Anthropic, *Claude Mythos Preview System Card* (San Francisco: Anthropic, 2026), https://www-cdn.anthropic.com/7624816413e9b4d2e3ba620c5a5e091b98b190a5.pdf.
[59] AI Security Institute, "Our Evaluation of Claude Mythos Preview's Cyber Capabilities," April 13, 2026, https://www.aisi.gov.uk/blog/our-evaluation-of-claude-mythos-previews-cyber-capabilities.

Another area of growing concern is biosecurity. Historically, biological risks were associated primarily with laboratory accidents, or the misuse of biological knowledge by malicious actors. AI may amplify these risks by making sophisticated biological expertise more accessible and by accelerating aspects of biological design and engineering. As a dual-use technology, AI may increase capabilities relevant to chemical, biological, radiological, and nuclear risks. Of particular concern is the convergence of AI and synthetic biology. Within design-build-test-learn cycles, AI systems can accelerate several stages of biological research, including the design and optimisation of biomolecules with potentially harmful applications, such as novel proteins and toxins, pathogen modification, or the creation of new biological agents (NASEM 2025).[61]

Recent examples illustrate both the promise and the risks of these developments. AI systems have already been used in the design of bacteriophages capable of targeting resistant bacterial strains, while models such as Evo 1 and Evo 2, the DNA foundation model from the Arc Institute, demonstrate sophisticated capabilities in genomic modelling and biological design across DNA, RNA, and proteins (Brixi et al. 2026).[62] Recognising these concerns, some developers have implemented safeguards. Evo 2, for example, although trained on both prokaryotic and eukaryotic genetic sequences, excluded viruses that infect eukaryotic organisms from its training data. Yet technical safeguards can potentially be circumvented, and debates continue over access restrictions for the most capable models.

The emergence of autonomous laboratories raises additional concerns. Systems capable of designing, conducting, and analysing experiments could dramatically accelerate scientific progress. Yet they could also be misused by malicious actors or deployed in ways that increase the risk of accidents. Scientific communities have historically developed mechanisms for managing dangerous technologies. Whether similar mechanisms will remain effective in an era of widely accessible biological knowledge and increasingly capable AI systems remains an open question.

---

[60] Tom Uren, "Mythos Fallout, U.S. Government Weighs AI Model Regulation," *Lawfare*, May 8, 2026, https://www.lawfaremedia.org/article/mythos-fallout--u.s.-government-weighs-ai-model-regulation.
[61] National Academies of Sciences, Engineering, and Medicine, The Age of AI in the Life Sciences: Benefits and Biosecurity Considerations (Washington, DC: National Academies Press, 2025), https://doi.org/10.17226/28868.
[62] Gary Brixi, Matthew G. Durrant, Jarome Ku, et al., "Genome Modelling and Design across All Domains of Life with Evo 2," *Nature* 652 (2026): 1349–1361, https://doi.org/10.1038/s41586-026-10176-5.

### 7.2. Long-Term and Existential Risks

The risks discussed so far arise primarily from the deployment of current or near-term AI systems. A different class of concerns focuses on future systems that may become substantially more autonomous and capable than those available today. These risks remain speculative, but they nonetheless shape contemporary debates about the long-term implications of advanced AI, particularly because current research continues to push rapidly toward more capable systems.

The most far-reaching concerns relate to the possibility of increasingly autonomous and self-improving AI systems. The capacity for self-improvement has long been regarded as one of the most important pathways toward superintelligence (Schmidhuber 2007).[63] Although contemporary systems do not yet exhibit unrestricted autonomous self-improvement, substantial research efforts are directed toward developing mechanisms that allow systems to improve their own performance with reduced human intervention.

Some evidence of this trajectory can already be seen in recent research. Results from the ARC Prize 2025 competition show the growing importance of self-improvement mechanisms. Systems in the competition used refinement loops that iteratively transform a program or model into a better-performing one by relying on feedback signals. Techniques such as recursive reasoning, self-improving evolutionary program synthesis, and test-time optimisation enabled even relatively small models, on the order of seven million parameters, to achieve competitive performance (Chollet et al. 2026)[64]. Evolutionary self-improvement frameworks typically combine an exploration phase, which generates multiple candidate solutions, with a verification phase that evaluates them and provides feedback. That feedback then guides iterative program refinement and optimisation.

Although such methods remain largely experimental, they demonstrate how self-improvement techniques can be generalised across architectures and potentially enhance AI systems' ability to solve complex scientific problems. At present, such systems do not constitute an immediate threat. However, sufficiently advanced forms of self-improvement could eventually contribute to the emergence of artificial superintelligence (ASI) whose capabilities exceed human understanding and control. In such scenarios, failures of alignment could result in severe forms of human disempowerment, especially if highly capable systems acquire advanced scientific, technological, social, or strategic knowledge. For this reason, many discussions of advanced AI risk focus on the development of robust mechanisms for oversight, alignment, and control.

---

[63] Jürgen Schmidhuber, "Gödel Machines: Fully Self-Referential Optimal Universal Self-Improvers," in *Artificial General Intelligence*, edited by Ben Goertzel and Cassio Pennachin (Berlin and Heidelberg: Springer, 2007), https://doi.org/10.1007/978-3-540-68677-4_7.
[64] Chollet "ARC Prize 2025: Technical Report".

## 8. Conclusion: The Future of Scientific Discovery

This article has presented a snapshot of the current state of AI in scientific research and discovery. The rapid diffusion of AI across the sciences has been enabled by advances within artificial intelligence itself, which have produced increasingly capable systems able to assist a growing range of research activities. Yet the achievements reviewed throughout this article should not be interpreted as evidence that progress will continue indefinitely, or even at its current pace.

Forecasting the development of AI remains notoriously imprecise. Current trends may continue, accelerate, slow, or plateau altogether. Predictions of imminent algorithmic breakthroughs coexist with recurring concerns about technical limitations, data constraints, economic pressures, regulatory interventions, and possible architectural bottlenecks. The history of AI has been marked both by periods of rapid growth and by periods of stagnation, including the so-called AI winters, as well as by repeated failures to predict the pace of progress accurately. Extrapolations from present trends should therefore be treated with caution. This applies no less to forecasts about the future role of AI in scientific discovery.

At present, despite substantial progress, AI systems have not produced paradigm-changing discoveries comparable to the most transformative ones in the history of science. They have contributed to important advances and have demonstrated significant abilities across a range of domains, but there is still a gap between solving well-defined scientific problems and generating discoveries that fundamentally change scientific views. This limitation should not be surprising. Historically, paradigm-changing discoveries have rarely emerged from isolated insights alone. They have usually depended on deep theoretical understanding, the synthesis of disparate domains of knowledge, and the gradual accumulation of theoretical and empirical insights. At the present stage of development, AI systems appear most effective when operating as powerful cognitive tools within existing research programmes designed and directed by human scientists. Their greatest successes have generally involved solving highly specific problems, which however already existed within established scientific frameworks.

At the same time, future systems may indeed acquire capacities and knowledge that differ not merely quantitatively but qualitatively from those observed today. Should that occur, scientific communities may encounter a novel challenge: AI-generated hypotheses, explanations, or theoretical frameworks may become increasingly difficult for humans to evaluate, interpret, or even comprehend. After all, the history of human science contains examples of premature discoveries that were initially ignored, rejected, or misunderstood because prevailing conceptual frameworks were not prepared to accommodate them as historical cases attest[65]. Similar difficulties could arise in relation to discoveries produced by advanced AI systems. Such outputs may appear implausible, unintelligible, or hallucinatory from the perspective of contemporary scientific paradigms, despite containing genuine insight. Human cognitive limitations could therefore become a significant obstacle to the assessment of increasingly sophisticated machine-generated knowledge.

Some indications of this possibility may already be emerging. Recent research suggests that AI systems can develop and employ abstract concepts that are not immediately transparent to human observers (Ramji et al. 2026).[66] Related work has documented the emergence of communication strategies and representational structures that differ from ordinary human language (Liu 2026).[67] Such developments raise important questions about the future interpretability of AI-generated scientific knowledge, but at the same time reveal potential barriers in the human-produced science.

A line of research seeks to explore precisely this possibility. For example, proponents of the "alien space of science" hypothesis argue that contemporary scientific communities may systematically overlook certain research directions because they lack the conceptual tools, intuitions, or methodological combinations necessary to recognise them (Artiles et al. 2026).[68] On that view, scientific progress may be constrained not only by technological limitations but also by the cognitive and institutional biases of human researchers. AI systems could then, at least in principle, help identify areas of science that remain effectively invisible within existing research traditions. At the same time, such a trajectory would inevitably increase the overall risks from AI systems present in scientific research.

For this reason, the future of AI in science cannot be evaluated solely in terms of innovation. Equally important are questions of ethics, governance, and control. The uncontrolled dissemination of scientific discoveries, the absence of accountability for consequences, such as the increasing militarisation of AI may undermine public trust in both science and technology. Scientists are likely to be judged not only by what they produce but also by the manner in which their discoveries are deployed, and governed.

---

[65] For example, Mendel's Laws of Inheritance and other.
[66] Keshav Ramji, Tahira Naseem, and Ramón Fernandez Astudillo, "Thinking Without Words: Efficient Latent Reasoning with Abstract Chain-of-Thought," arXiv, 2026, https://doi.org/10.48550/arXiv.2604.22709.
[67] Hung Ming Liu, "AI Mother Tongue: Self-Emergent Communication in MARL via Endogenous Symbol Systems," arXiv, 2025, https://doi.org/10.48550/arXiv.2507.10566.
[68] Alejandro H. Artiles, Martin Weiss, Levin Brinkmann, et al., "The Alien Space of Science: Sampling Coherent but Cognitively Unavailable Research Directions," arXiv, 2026, https://doi.org/10.48550/arXiv.2603.01092

**Disclosure**

The author used GPT-5.5 during the preparation of this manuscript to improve language and readability and to assist with literature exploration. All AI-assisted output was critically reviewed, verified, and revised by the author, who assumes full responsibility for the content of the manuscript.

**Funding**

This work has been funded by a grant from the Programme Johannes Amos Comenius under the Ministry of Education, Youth and Sports of the Czech Republic, CZ.02.01.01/00/23_025/0008711.

**Bibliography**

Abramson, Josh, Jonas Adler, Jack Dunger, et al. "Accurate Structure Prediction of Biomolecular Interactions with AlphaFold 3." Nature 630 (2024): 493–500. https://doi.org/10.1038/s41586-024-07487-w.

Agüera y Arcas, Blaise. "Do Large Language Models Understand Us?" Daedalus 151, no. 2 (2022): 183–197. https://doi.org/10.1162/DAED_a_01909.

Alper, J., M. Barany, A. Chavarri Villarello, S. Dahmen, W. Dean, K. Ganapathy, M. Harris, D. Holmes, M. Jamnik, S. Kelk, B. Kra, U. Martin, B. Naskręcki, R. Ochigame, J. Portegies, and J. Schmitt. Leiden Declaration on Artificial Intelligence and Mathematics. Zenodo, 2026. https://doi.org/10.5281/zenodo.20302944

Anthropic. Claude Mythos Preview System Card. San Francisco: Anthropic, 2026. https://www-cdn.anthropic.com/7624816413e9b4d2e3ba620c5a5e091b98b190a5.pdf.

Artiles, Alejandro H., Martin Weiss, Levin Brinkmann, Iyad Rahwan, Bernhard Schölkopf, Christopher Pal, Hugo Larochelle, Anirudh Goyal, and Nasim Rahaman. 2026. "The Alien Space of Science: Sampling Coherent but Cognitively Unavailable Research Directions." arXiv. https://doi.org/10.48550/arXiv.2603.01092

Bacon, Francis. Essayes. Religious Meditations. Places of Perswasion and Disswasion. Seene and Allowed. London: Printed [by John Windet] for Humfrey Hooper, 1597. Accessed June 12, 2026. Internet Archive. https://archive.org/details/bim_early-english-books-1475-1640_essayes-religious-medit_bacon-francis-viscount_1597.

Costello, Thomas H., Gordon Pennycook, and David G. Rand. "Durably Reducing Conspiracy Beliefs through Dialogues with AI." Science 385, no. 6714 (2024): eadq1814. https://doi.org/10.1126/science.adq1814

Batatia, Ilyes, Philipp Benner, Yuan Chiang, Alin M. Elena, Dávid P. Kovács, Janosh Riebesell, Xavier R. Advincula, Mark Asta, Matthew Avaylon, William J. Baldwin, et al. "A Foundation Model for Atomistic Materials Chemistry." The Journal of Chemical Physics 163, no. 18 (2025): 184110. https://doi.org/10.1063/5.0297006.

Bender, Emily M., Timnit Gebru, Angelina McMillan-Major, and Shmargaret Shmitchell. "On the Dangers of Stochastic Parrots: Can Language Models Be Too Big?" In Proceedings of the 2021 ACM Conference on Fairness, Accountability, and Transparency, 610–623. New York: Association for Computing Machinery, 2021. https://doi.org/10.1145/3442188.3445922.

Binksmith, Adam et. al. "A New Moore's Law for AI Agents." AI Digest, March 24, 2026. Accessed June 12, 2026. https://theaidigest.org/time-horizons.

Binz, Marcel, Elif Akata, Matthias Bethge, et al. "A Foundation Model to Predict and Capture Human Cognition." Nature 644 (2025): 1002–1009. https://doi.org/10.1038/s41586-025-09215-4.

Brixi, Gary, Matthew G. Durrant, Jarome Ku, et al. “Genome Modelling and Design across All Domains of Life with Evo 2.” Nature 652 (2026): 1349–1361. https://doi.org/10.1038/s41586-026-10176-5.

Center for AI Safety, Scale AI, and HLE Contributors Consortium. “A Benchmark of Expert-Level Academic Questions to Assess AI Capabilities.” Nature 649 (2026): 1139–1146. https://doi.org/10.1038/s41586-025-09962-4.

Center for AI Safety. Humanity's Last Exam. Accessed June 12, 2026. https://lastexam.ai/.

Chollet, François, Mike Knoop, Gregory Kamradt, and Bryan Landers. “ARC Prize 2025: Technical Report.” arXiv, 2026. https://doi.org/10.48550/arXiv.2601.10904.

Dawid, Anna, and Yann LeCun. “Introduction to Latent Variable Energy-Based Models: A Path Towards Autonomous Machine Intelligence.” arXiv, 2023. https://doi.org/10.48550/arXiv.2306.02572.

Ding, Liangping, Cornelia Lawson, and Philip Shapira. “Rise of Generative Artificial Intelligence in Science.” Scientometrics 130 (2025): 5093–5114. https://doi.org/10.1007/s11192-025-05413-z.

Ghareeb, Ali Essam, Benjamin Chang, Ludovico Mitchener, et al. “A Multi-Agent System for Automating Scientific Discovery.” Nature (2026). https://doi.org/10.1038/s41586-026-10652-y.

Glickman, Moshe, and Tali Sharot. “How Human–AI Feedback Loops Alter Human Perceptual, Emotional and Social Judgements.” Nature Human Behaviour 9 (2025): 345–359. https://doi.org/10.1038/s41562-024-02077-2.

Gottweis, Juraj, Wei-Hung Weng, Alexander Daryin et al. “Accelerating Scientific Discovery with Co-Scientist.” Nature (2026). https://doi.org/10.1038/s41586-026-10644-y.

Guevara, Alfredo, Alexandru Lupsasca, David Skinner, Andrew Strominger, and Kevin Weil. “Single-Minus Gluon Tree Amplitudes Are Nonzero.” arXiv, 2026. https://doi.org/10.48550/arXiv.2602.12176.

Ju, Haocheng, Guoxiong Gao, Jiedong Jiang, Bin Wu, Zeming Sun, Leheng Chen, Yutong Wang, Yuefeng Wang, Zichen Wang, Wanyi He, et al. “Automated Conjecture Resolution with Formal Verification.” arXiv, 2026. https://doi.org/10.48550/arXiv.2604.03789.

Jumper, John, Richard Evans, Alexander Pritzel, et al. “Highly Accurate Protein Structure Prediction with AlphaFold.” Nature 596 (2021): 583–589. https://doi.org/10.1038/s41586-021-03819-2.

Kantrowitz, Alex. “Google DeepMind CEO Demis Hassabis: AI's Next Breakthroughs, AGI Timeline, Google's AI Glasses Bet.” Big Technology Podcast. YouTube video, premiered January 23, 2026. https://www.youtube.com/watch?v=bgBfobN2A7A.

Ke, Zixuan, Fangkai Jiao, Yifei Ming, Xuan-Phi Nguyen, Austin Xu, Do Xuan Long, Minzhi Li, Chengwei Qin, Peifeng Wang, Silvio Savarese, Caiming Xiong, and Shafiq Joty. “A Survey of Frontiers in LLM Reasoning: Inference Scaling, Learning to Reason, and Agentic Systems.” arXiv, 2025. https://doi.org/10.48550/arXiv.2504.09037.

Knuth, Donald E. “Claude Cycles.” PDF manuscript. Accessed June 11, 2026. https://www-cs-faculty.stanford.edu/~knuth/papers/claude-cycles.pdf.

Ko, Kuang-Ting, Frank Lennartz, David Mekhaiel, Bora Guloglu, Arianna Marini, Danielle J. Deuker, Carole A. Long et al. "Structure of the Malaria Vaccine Candidate Pfs48/45 and Its Recognition by Transmission Blocking Antibodies." Nature Communications 13, no. 1 (2022): 5603. https://doi.org/10.1038/s41467-022-33379-6.

Kwa, Thomas, Ben West, Joel Becker, Amy Deng, Katharyn Garcia, Max Hasin, Sami Jawhar, Megan Kinniment, Nate Rush, Sydney von Arx, et al. "Measuring AI Ability to Complete Long Tasks." arXiv, 2025. https://doi.org/10.48550/arXiv.2503.14499.

Langley, Pat, Herbert A. Simon, and Gary L. Bradshaw. 1987. "Heuristics for Empirical Discovery." In Computational Models of Learning, edited by Leonard Bolc, 21–54. Berlin and Heidelberg: Springer-Verlag. https://doi.org/10.1007/978-3-642-82742-6_2.

Lee, Seungpil, Woochang Sim, Donghyeon Shin, Wongyu Seo, Jiwon Park, Seokki Lee, Sanha Hwang, Sejin Kim, and Sundong Kim. "Reasoning Abilities of Large Language Models: In-Depth Analysis on the Abstraction and Reasoning Corpus." ACM Transactions on Intelligent Systems and Technology 16, no. 6 (2025): Article 137, 1–52. https://doi.org/10.1145/3712701.

Liang, Weixin, Yuhui Zhang, Zhengxuan Wu, et al. "Quantifying Large Language Model Usage in Scientific Papers." Nature Human Behaviour 9 (2025): 2599–2609. https://doi.org/10.1038/s41562-025-02273-8.

Liu, Hung Ming. "AI Mother Tongue: Self-Emergent Communication in MARL via Endogenous Symbol Systems." arXiv, 2025. https://doi.org/10.48550/arXiv.2507.10566.

Lowe, Derek. "Evaluating 'Co-Scientist', a New AI Science System." In the Pipeline (blog). Science, May 27, 2026. https://www.science.org/content/blog-post/evaluating-co-scientist-new-ai-science-system.

Lu, Chris, Cong Lu, Robert Tjarko Lange, et al. "Towards End-to-End Automation of AI Research." Nature 651 (2026): 914–919. https://doi.org/10.1038/s41586-026-10265-5.

Lu, Chris, Cong Lu, Robert Tjarko Lange, Jakob Foerster, Jeff Clune, and David Ha. "The AI Scientist: Towards Fully Automated Open-Ended Scientific Discovery." arXiv, 2024. https://doi.org/10.48550/arXiv.2408.06292.

National Academies of Sciences, Engineering, and Medicine. The Age of AI in the Life Sciences: Benefits and Biosecurity Considerations. Washington, DC: National Academies Press, 2025. https://doi.org/10.17226/28868.

Nature. Search results page. Accessed June 11, 2026. https://www.nature.com/.

Novikov, Alexander, Ngân Vũ, Marvin Eisenberger, Emilien Dupont, Po-Sen Huang, Adam Zsolt Wagner, Sergey Shirobokov, Borislav Kozlovskii, Francisco J. R. Ruiz, Abbas Mehrabian, et al. "AlphaEvolve: A Coding Agent for Scientific and Algorithmic Discovery." arXiv, 2025. https://doi.org/10.48550/arXiv.2506.13131.

OpenAI. "OpenAI Model Disproves Discrete Geometry Conjecture." Accessed June 11, 2026. https://openai.com/cs-CZ/index/model-disproves-discrete-geometry-conjecture/.

Ramji, Keshav, Tahira Naseem, and Ramón Fernandez Astudillo. "Thinking Without Words: Efficient Latent Reasoning with Abstract Chain-of-Thought." arXiv, 2026. https://doi.org/10.48550/arXiv.2604.22709.

Romera-Paredes, Bernardino, Mohammadamin Barekatain, Alexander Novikov, et al. "Mathematical Discoveries from Program Search with Large Language Models." Nature 625 (2024): 468–475. https://doi.org/10.1038/s41586-023-06924-6.

Sajadieh, Sha, Loredana Fattorini, Raymond Perrault, Yolanda Gil, Vanessa Parli, Lapo Santarlasci, Juan Pava, Nestor Maslej, Russ Altman, and Erik Brynjolfsson, et al. The AI Index 2026 Annual Report. Stanford, CA: AI Index Steering Committee, Institute for Human-Centered AI, Stanford University, 2026. https://hai.stanford.edu/assets/files/ai_index_report_2026.pdf.

Salvi, Francesco, Manoel Horta Ribeiro, Riccardo Gallotti, et al. "On the Conversational Persuasiveness of GPT-4." Nature Human Behaviour 9 (2025): 1645–1653. https://doi.org/10.1038/s41562-025-02194-6.

Samadder, Jewel. "Ongoing Phase 1b/2 Trial of the Allosteric MEK1/2 Inhibitor REC-4881 as Monotherapy in Familial Adenomatous Polyposis (FAP): Preliminary Safety and Efficacy Data." Presentation, May 4, 2025. Recursion Pharmaceuticals. PDF. https://ir.recursion.com/static-files/cef86a12-64aa-4ff9-9074-2f81fb3efd17.

Schmidhuber, Jürgen. "Gödel Machines: Fully Self-Referential Optimal Universal Self-Improvers." In Artificial General Intelligence, edited by Ben Goertzel and Cassio Pennachin. Berlin and Heidelberg: Springer, 2007. https://doi.org/10.1007/978-3-540-68677-4_7.

Schoenegger, Philipp, Spencer Greenberg, Alexander Grishin, et al. "AI Can Outperform Humans in Predicting Correlations Between Personality Items." Communications Psychology 3 (2025): 23. https://doi.org/10.1038/s44271-025-00205-w.

Science Friday. "How AlphaFold Has Changed Biology Research, 5 Years On." YouTube video. Accessed June 11, 2026. https://www.youtube.com/watch?v=n1rr36vHG8A.

Soares, Eduardo, Emilio Vital Brazil, Victor Shirasuna, et al. "An Open-Source Family of Large Encoder-Decoder Foundation Models for Chemistry." Communications Chemistry 8 (2025): 193. https://doi.org/10.1038/s42004-025-01585-0.

Song, Tao, Man Luo, Xiaolong Zhang, LinJiang Chen, Yan Huang, Jiaqi Cao, Qing Zhu, Daobin Liu, Baicheng Zhang, Gang Zou, et al. "A Multiagent-Driven Robotic AI Chemist Enabling Autonomous Chemical Research On Demand." Journal of the American Chemical Society 147, no. 15 (2025): 12534–12545. https://doi.org/10.1021/jacs.4c17738.

Song, Zhangde, Jieyu Lu, Yuanqi Du, Botao Yu, Thomas M. Pruyn, Yue Huang, Kehan Guo, Xiuzhe Luo, Yuanhao Qu, Yi Qu, et al. "Evaluating Large Language Models in Scientific Discovery." arXiv preprint arXiv:2512.15567. Version 2, May 8, 2026. https://doi.org/10.48550/arXiv.2512.15567.

Stokel-Walker, Chris. "Scientists Invented a Fake Disease. AI Told People It Was Real." Nature 652, no. 8110 (2026): 559–561. https://doi.org/10.1038/d41586-026-01100-y.

Strachan, James W. A., Dalila Albergo, Giulia Borghini, et al. “Testing Theory of Mind in Large Language Models and Humans.” Nature Human Behaviour 8 (2024): 1285–1295. https://doi.org/10.1038/s41562-024-01882-z.

Topaz, Maxim, Nir Roguin, Palavi Gupta, et al. “Fabricated Citations: An Audit across 2.5 Million Biomedical Papers.” The Lancet 407 (2026): 1779–1781.

Tsoukalas, George, Anton Kovsharov, Sergey Shirobokov, et al. “Advancing Mathematics Research with AI-Driven Formal Proof Search.” arXiv, 2026. https://doi.org/10.48550/arXiv.2605.22763.

UK AI Security Institute. “Our Evaluation of Claude Mythos Preview’s Cyber Capabilities.” April 13, 2026. https://www.aisi.gov.uk/blog/our-evaluation-of-claude-mythos-previews-cyber-capabilities.

Uren, Tom. “Mythos Fallout, U.S. Government Weighs AI Model Regulation.” Lawfare, May 8, 2026. https://www.lawfaremedia.org/article/mythos-fallout--u.s.-government-weighs-ai-model-regulation.

Wenkel, Frederik, Wilson Tu, Cassandra Masschelein, et al. “TxPert: Using Multiple Knowledge Graphs for Prediction of Transcriptomic Perturbation Effects.” Nature Biotechnology (2026). https://doi.org/10.1038/s41587-026-03113-4.

Wiesner, Florian, Zoë J. Gray, Matthias Wessling, and Stephen Baek. “Towards a Physics Foundation Model.” arXiv, 2025. https://arxiv.org/abs/2509.13805.